\documentclass[letterpaper, 10 pt,conference]{ieeeconf}
\IEEEoverridecommandlockouts                              

\usepackage{cite}
\usepackage{amsmath,amssymb,amsfonts}
\usepackage{graphicx}
\usepackage{textcomp}
\usepackage{xcolor}
\usepackage{algorithm}
\usepackage{algcompatible}
\usepackage{microtype}
\usepackage{array}
\usepackage{tikz,}
\usepackage{graphicx}
\usepackage[utf8]{inputenc}
\usepackage[font=footnotesize]{subfig}
\usetikzlibrary{positioning}
\usetikzlibrary{arrows}
\IEEEoverridecommandlockouts   
\usepackage{caption}
\newcolumntype{P}[1]{>{\centering\arraybackslash}p{#1}}
\newcolumntype{M}[1]{>{\centering\arraybackslash}m{#1}}

\makeatletter
\newcommand{\removelatexerror}{\let\@latex@error\@gobble}
\makeatother

\begin{document}

\title{\LARGE \bf
The Braess’s Paradox in Dynamic Traffic
}

\author{Dingyi Zhuang$^{1*}$, Yuzhu Huang$^{1*}$, Vindula Jayawardana$^{23}$, Jinhua Zhao$^{1}$, Dajiang Suo$^{24}$, Cathy Wu$^{12}$  
\thanks{*Dingyi Zhuang and Yuzhu Huang have contributed equally.}
\thanks{$^{1}$ Department of Civil and Environmental Engineering, Massachusetts Institute of Technology {\tt\small dingyi@mit.edu}; {\tt\small yuzhuh@mit.edu}; {\tt\small jinhua@mit.edu}}%
\thanks{$^{2}$ Laboratory for Information and Decision Systems, Massachusetts Institute of Technology {\tt\small vindula@mit.edu}; {\tt\small cathywu@mit.edu}}
\thanks{$^{3}$ Department of Electrical Engineering and Computer Science, Massachusetts Institute of Technology}
\thanks{$^{4}$ Department of Mechanical Engineering, Massachusetts Institute of Technology {\tt\small  djsuo@mit.edu}}%
}

\maketitle



\begin{abstract}
The Braess's Paradox (BP) is the observation that adding one or more roads to the existing road network will counter-intuitively increase traffic congestion and slow down the overall traffic flow. Previously, the existence of the BP is modeled using the static traffic assignment model, which solves for the user equilibrium subject to network flow conservation to find the equilibrium state and distributes all vehicles instantaneously. Such approach neglects the dynamic nature of real-world traffic, including vehicle behaviors and the interaction between vehicles and the infrastructure. As such, this article proposes a \textit{dynamic} traffic network model and empirically validates the existence of the BP under dynamic traffic. In particular, we use microsimulation environment to study the impacts of an added path on a 2-by-3 all stop-controlled grid network. We explore how the network flow, vehicle travel time, and network capacity respond, as well as when the BP will occur. We find that adding a path to the network does not increase the overall network output flow and worsens it when the added path is more attractive. We observe that traffic flow on the added path increases first, but decreases after the demand level reaches an inflection point. The added path worsens the average travel time of all vehicles and does not provide additional operational capacity to the network.
\end{abstract}

\section{Introduction}
\label{intro}
\subsection{Background}
Dietrich Braess discovered that adding a new road to an existing road network, although increasing capacity, could exacerbate overall congestion. He argued that if each vehicle is making self-interested routing decisions, the shortest path would be frequently selected, leading to impedance on the shortcut \cite{steinberg1983prevalence,frank1981braess}. A classical approach to formulating this problem usually involves using road cost function related to the number of vehicles on each road to solve for the traffic equilibrium equations and demonstrate the BP \cite{nagurney2021braess}. 

\begin{figure}[t]
     \centering
     \subfloat[The four-edge network\label{1a}]{%
         \begin{tikzpicture}[>=triangle 45,font=\sffamily]
            \node (X) at (0,0) {A};
            \node (Y) [below left=1cm and 1cm of X]  {C};
            \node (Z) [below right=1cm and 1cm of X] {D};
            \node (U) [below left=1cm and 1cm of Z]  {B};
            \draw [semithick,->] (X) -- (Y) node [midway,above,sloped] {10N};
            \draw [semithick,->] (X) -- (Z) node [midway,above,sloped] {N+50};
            \draw [semithick,->] (Y) -- (U) node [midway,below,sloped] {N+50};
            \draw [semithick,->] (Z) -- (U) node [midway,below,sloped] {10N};
        \end{tikzpicture}}
        \hspace{15mm}
     \subfloat[The five-edge network\label{1b}]{%
         \begin{tikzpicture}[>=triangle 45,font=\sffamily]
            \node (X) at (0,0) {A};
            \node (Y) [below left=1cm and 1cm of X]  {C};
            \node (Z) [below right=1cm and 1cm of X] {D};
            \node (U) [below left=1cm and 1cm of Z]  {B};
            \draw [semithick,->] (Y) -- (Z) node [midway,above,sloped] {N+10};
            \draw [semithick,->] (X) -- (Y) node [midway,above,sloped] {10N};
            \draw [semithick,->] (X) -- (Z) node [midway,above,sloped] {N+50};
            \draw [semithick,->] (Y) -- (U) node [midway,below,sloped] {N+50};
            \draw [semithick,->] (Z) -- (U) node [midway,below,sloped] {10N};
        \end{tikzpicture}}
     \caption{The classical Braess's Paradox example with static traffic assignment, where $N$ represents the number of vehicles on the edge. If vehicles reach equilibrium, adding a path (i.e. connection CD in b) will make the traffic worse.}
     \label{fig:bp}
\end{figure}
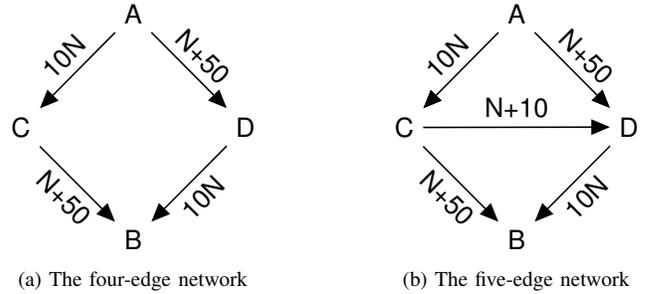

To see how to solve the equilibrium for the BP, consider a classical static traffic assignment example. As shown in Fig. \ref{fig:bp}a, the diamond-shaped network, suppose there are six cars that wish to travel from A to B via routes \textit{ACB} or \textit{ADB}. The travel cost (usually travel time) for each road segment is marked beside the edge, where $N$ is the number of vehicles on the same road segment. Intuitively, if there are more vehicles on the same road, traffic tends to congest and travel time is consequently larger. Since the cost function of each route is the same for routes \textit{ACB} and \textit{ADB} (if \textit{CD} is not connected), the demand is evenly distributed because using one route is just as good as using the other. The user equilibrium reaches when three cars use route ACB and the other three choose ADB with travel cost $C_{ACB} = 30+53 = 83 = C_{ADB}$. Now, suppose a new shortcut \textit{CD} connecting the two routes is constructed, as shown in Fig. \ref{fig:bp}b. The four-edge network is reformed to the five-edge one. Vehicles will shift to the shortcut until a new user equilibrium is reached when $C_{ACDB} = C_{ACB} = C_{ADB}$, when two vehicles choose each route and the cost is 92 for all, larger than the previous case without the shortcut. From this example, we could see that adding a new road to the network could in fact increase the travel time for everybody, which is the Braess's Paradox.

\subsection{Motivation}
The application of static traffic assignment on the four-edge and five-edge networks is widely discussed in previous literature on Braess's Paradox \cite{bittihn2021effect,fu2013analysis}. These works analyze the BP from the network science perspective. For the convenience of optimization, vehicle behaviors and the routing strategy are usually simplified. Vehicles are homogeneous and their route choices are assigned instantaneously after solving the flow conservation and user equilibrium. Wolf et al. \cite{wolf1996traffic} proposed the TASEP framework to study the BP, but only discuss the random and fixed routing of vehicles. The lack of consideration of vehicle behaviors makes the results less empirical in real-world scenario. Colombo, et al. modeled the BP from the microscopic vehicular traffic, but still requires a lot of constraints and assumptions to solve the Nash equilibrium \cite{colombo2020microscopic}. Recently, Belov, Aleksandr, et al. \cite{belov2021microsimulation} studied the price of anarchy from the Braess network using the microsimulation. The price of anarchy is the index between the user equilibrium and the system optimal. They have also discussed the network parameters and their impacts on the price of anarchy and related it with the BP. To be differentiated, our work has designed a estimated travel time based routing strategy and discuss directly the added path's impact on the network and the BP.

Moreover, under a static traffic assignment model, another crucial hidden assumption is that vehicles are assigned instantaneously. This assumption is impractical as the-real world vehicles' behavior dynamically changes with the surrounding environment. Moreover, the junction effects and the interactions of human drivers are not fully studied. Therefore, we wish to examine whether BP exists in the dynamic setting. Our approach is to mimic the vehicle behaviors and analyze the network performance from the microscopic perspective. In this paper, we use the Flow project and SUMO \cite{wu2021flow,vinitsky2018benchmarks} as the microsimulation environment to design minimalistic scenarios and explore the characteristics of BP in the dynamic setting. Our main contributions include:

\begin{itemize}
    \item We extend the static traffic assignment problem into the microsimulation context to model the BP
    \item Under dynamic traffic, we empirically explore the impacts of an added path on the network flow, network capacity, and vehicle travel time, and confirm the existence of BP in dynamic traffic
\end{itemize}

To our best knowledge, we are the first to analyze the BP from the dynamic microsimulation setting.


\section{A Dynamic Braess's Paradox Model}

\subsection{Preliminaries}
Fig. \ref{fig:bp} example shows the idea of static traffic assignment model. We briefly review its formulation and relevant extension in the dynamic traffic assignment context, before extending it to a dynamic microsimulation one \cite{fu2013analysis}. It assumes that the traffic network contains a set $\mathcal{S}$ of origin-destination pairs with $n_\mathcal{S}$ elements, the edge set $\mathcal{L}$ and the path route set $\mathcal{P}$ \cite{fu2013analysis}. Shown in Fig. \ref{fig:bp} example, each route $p\in \mathcal{P}$ has cost function $C_p$, whose mathematical formulation of the user equilibrium is given as \cite{dafermos1969traffic,smith1979existence,ng2012dynamic,meng2012computational}:
\begin{equation}
    \begin{aligned}
        C_p(x_p^*(t)) &= \lambda_s(t), \text{  if }x_p^*(t) >0\\
        C_p(x_p^*(t)) &\geq \lambda_s(t), \text{  if }x_p^*(t) =0
    \end{aligned},
\end{equation}
where $C_p(\cdot)$ defines the user (travel) cost on path $p$, $\lambda_{s}(t)$ is the minimal path costs at time $t$, and $x_p^*(t)$ is the path flow assignment that fulfills flow conservation. In static assignment, $C_p$ is fixed while in dynamic assignment, $C_p(t)$ varies at different time. To help keep the notation consistency when it is extended to the dynamic traffic assignment, we keep all the variable dependent on time. The flow conservation is in the form as:
\begin{equation}
    \begin{aligned}
        f_e(t) = \sum_{p\in \mathcal{P}} x_p(t)\delta_{ep}
    \end{aligned},
\end{equation}
where $f_e(t)$ is the flow of the edge $e$ at time $t$, $x_p(t)$ is the flow on route $p$, and $\delta_{ep}$ is a binary indicator for whether edge $e$ is included in route $p$. The traffic demand $d_s(t)$ at time $t$ must also follow flow conservation constraints:
\begin{equation}
    \begin{aligned}
        \sum_{p\in \mathcal{P}} x_p(t) = d_s(t), \quad \forall s\in \mathcal{S}, \forall t
    \end{aligned}.
\end{equation}
Similarly, denote the user cost on the edge $e$ as $C_e(\cdot)$. The user costs on paths are defined as:
\begin{equation}
    \begin{aligned}
        C_p(t) = \sum_{a\in \mathcal{L}} C_e(t) \delta_{ep}, \quad \forall p\in \mathcal{P}, \forall t
    \end{aligned}.
\end{equation}
Dynamic traffic assignment takes as $C_p(t), C_e(t), d_s(t)$ as inputs and distributes vehicles by computing the flow conservation, minimizing each user's (travel) costs $\sum_{p\in \mathcal{P}}C_p(t)$, and ensuring user travel costs reach equilibrium at every time step. As shown in the Fig. \ref{fig:bp} example, $\sum_{p\in \mathcal{P}}C_p$ is also used to show the existence of Braess's Paradox in the static setting. The design of $C_p$ is usually simplified to depend on only the flow on the edge. Solving the user equilibrium takes longer when the traffic network size is large. We try to replace it with dynamic microsimulation, where we can capture the microscopic vehicle behaviors and understand how equilibrium and the BP naturally form.

\subsection{Problem Formulation}
To mimic the real-world driving behaviors, we take a microsimulation perspective. We wish to replace the flow conservation and user equilibrium optimization with the designed routing strategy for each vehicle. Denote the driver set $\mathcal{U} = \{d_s(t)\}$ that contains the demand at time $t$. When choosing the route, we consider each vehicle $u \in \mathcal{U}$ has potential travel cost function $C_u$ correlated to designed function $g(\cdot)$ of route flow, travel time, edge information:
\begin{equation}
    C_u(t) = g(x_p(t),e,p), \quad\forall e\in \mathcal{L},\forall p\in \mathcal{P}.
\end{equation}
We formulate the microscopic traffic assignment model by equalizing $\sum_{u\in\mathcal{U}} C_u$ instead of $\sum_{p\in\mathcal{P}} C_p$ by solving user equilibrium. The introduction of $C_u$ provides us with the microscopic leverage to discuss the important factors that will influence the vehicle routing decisions, as well as the potential impacts on the whole network.

\label{sec:design}
\subsection{Network Design}
The classical diamond-shaped networks in Fig. \ref{fig:bp} have some distinct characteristics. Traffic enters from node A and leaves at node B. Before a new path is added, each route from node A to node B (A$\rightarrow$C$\rightarrow$B, or A$\rightarrow$D$\rightarrow$B, shown in Fig. \ref{1a}) has the same overall cost function; after the new path is added, the third route (A$\rightarrow$C$\rightarrow$D$\rightarrow$B, shown in Fig. \ref{1b}) has a different cost function that makes it more preferable than the original routes.

\begin{figure} 
    \centering
    \subfloat[Gird network with 2 routes (the baseline network/scenario).
    \label{fig:diamond_ori}]{%
   \includegraphics[width=0.48\linewidth]{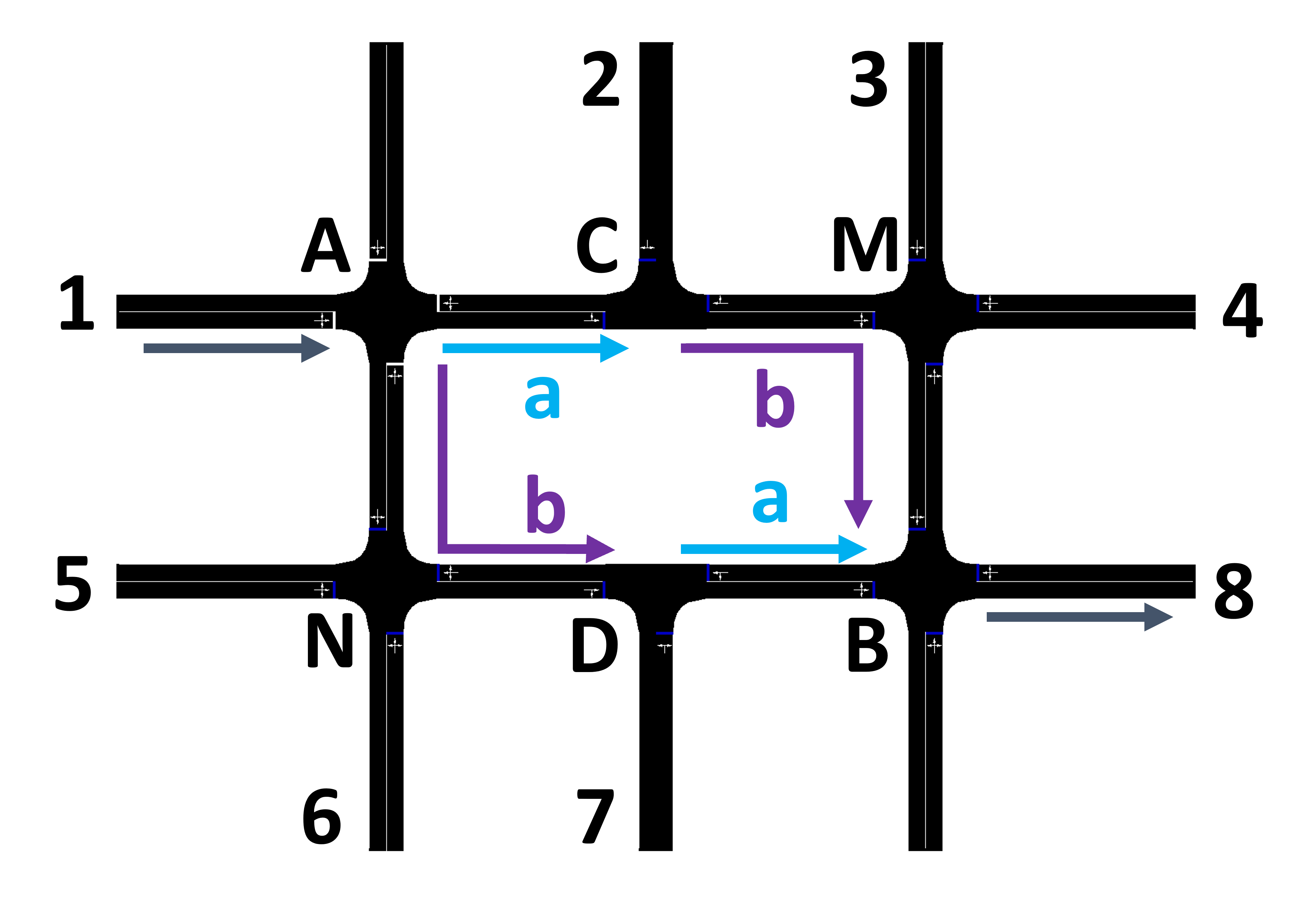}}
   \hfill
   \subfloat[Grid network with an added path (the add-path network/scenario).
    \label{fig:diamond_ap}]{%
   \includegraphics[width=0.48\linewidth]{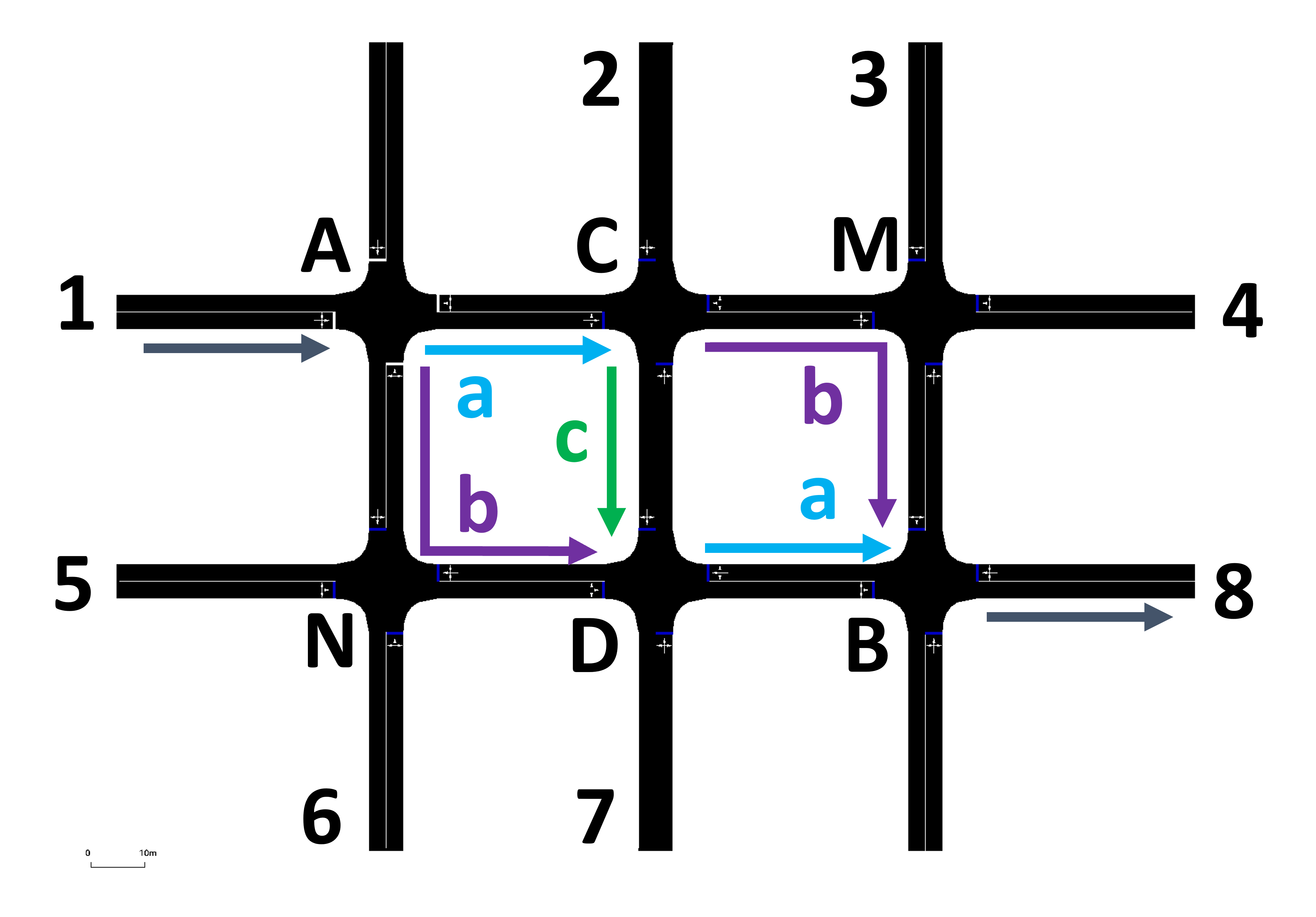}}
  \caption{A 2-by-3 grid network resembling the classical diamond network.}
  \label{fig:diamond} 
\end{figure}

We designed a grid network to resemble the classic diamond network, as shown in Fig. \ref{fig:diamond}. There are 6 intersections, all controlled with all-way stops except for intersection A. Arrows denote the directions of traffic flow. In the basic design, shown in Fig. \ref{fig:diamond_ori}, two routes are available, namely A$\rightarrow$C$\rightarrow$M$\rightarrow$B and A$\rightarrow$N$\rightarrow$D$\rightarrow$B. With an additional path added, as shown in Fig. \ref{fig:diamond_ap}, a third route A$\rightarrow$C$\rightarrow$D$\rightarrow$B becomes available.

\subsection{Cost Function $C_u$}

Unlike the classical diamond network that has a pre-defined linear cost function $C_p$ and $C_e$ , the dynamic microsimulation setting has a dynamic cost $C_u$ changing in real time. When a vehicle is choosing which route to take, it predicts the travel time of all possible routes, and chooses the one with the lowest predicted travel time.

In microsimulation, each vehicle operates with its own controller that determines the vehicle kinematics conditioned on its surrounding traffic, for example, following the IDM model that is described in Equation \ref{eq:idm}: 
\begin{equation}\label{eq:idm}
    s^*(v, \Delta v)=s_0+max[0,(vT+\frac{v\Delta v}{2\sqrt{ab}})],
\end{equation}
where the desired distance $s^*$ is the sum of minimum spacing $s_0$ and the acceleration/braking distance that is determined by the current vehicle speed $v$, desired headway $T$, relative speed $\Delta v$, maximum vehicle acceleration $a$, and a comfortable braking deceleration $b$ \cite{treiber2000congested}. Each vehicle routes itself based on the real-time prediction of travel times of all potential routes. This is equivalent to vehicles equipped with a real-time traveler information system (e.g. Google Maps).

A typical method could use historical travel times to estimate the future. However, historical travel time is not always representative of future travel times when the edge length is long or congestion builds up quickly, which is the case in several of the scenarios in this study. Therefore, the method here is to use the position and speed of all vehicles to estimate the travel time of each edge, then summing them up to get the estimated time of the route. The idea is to estimate the travel time using only readily-available information (speed, acceleration) without solving ODEs. Each vehicle makes the routing decision based on the estimation of the traffic condition on each route.

\begin{figure}[htbp]
    \centering
    {%
   \includegraphics[width=0.95\linewidth]{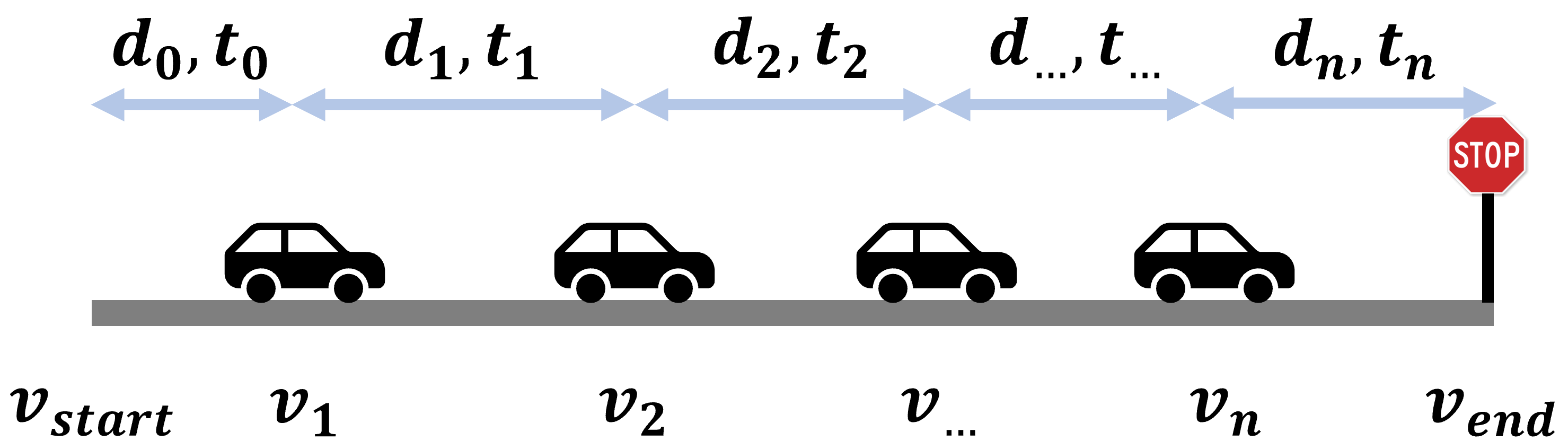}}
  \caption{Example of a stop-controlled edge with n vehicles.}
  \label{fig:tt_calc} 
\end{figure}

We estimate travel time of an edge as illustrated in Fig. \ref{fig:tt_calc}, where $d_0$ is the distance between the beginning of the edge and vehicle $1$ (the last vehicle on the edge), $d_i$ is the distance between vehicle $i$ and $i+1$, and $d_n$ is the distance between vehicle $n$ (the front vehicle on the edge) and the stop sign.

\noindent Therefore, the total length of the edge should be:
\begin{equation}
    L_{edge} = \sum_{i=0}^{n}d_i,
\end{equation}
Estimated travel time of an edge:
\begin{equation}
    C_e(edge) = 
        \begin{cases}
            \frac{L_{edge}}{v_{max}} &\text{if edge is empty}\\
            \sum_{i=0}^{n}t_i &\text{otherwise}
        \end{cases},
\end{equation}
where $v_{max}$ is the speed limit of the edge, $t_0$ is the estimated time for a vehicle to travel from the beginning of the edge to the position of vehicle $1$, $t_i$ is the estimated travel time for vehicle $i$ to travel to the position of vehicle $i+1$ and $t_n$ is the estimated time for vehicle $n$ to travel to the stop sign.

The decomposed travel times $t_i (i\in\{0, 1, ..., n\})$ as labeled in Fig. \ref{fig:tt_calc} are calculated using vehicle kinematics equations and assuming that vehicles have constant acceleration/deceleration rates. The travel time between two points (e.g. ego vehicle and a target vehicle) is calculated as described in Algorithm \ref{alg:est_tt}.
\begin{algorithm}
\caption{Estimate Travel Time from Current Position to Target Position ($\bf{est\_tt}$)}
\label{alg:est_tt}
\textbf{Given}: speed at current position $v_{ego}$, current acceleration $a_{ego}$, speed at target position $v_{target}$, maximum possible speed on edge $v_{max}$, constant acceleration $a$, distance between current and target positions $d$

\begin{algorithmic}[1]
\IF{$a_{ego}$ is small}
    \STATE assume that vehicle travels at constant speed until it has to accelerate to reach $v_{target}$
    \STATE $v_{final}\gets$ realized speed at target position,
    \STATE \quad\quad\quad\quad $=min(\sqrt{v_{ego}^2+2ad},v_{target})$
    \STATE $d_{accel}\gets$ acceleration distance $|\frac{v_{ego}^2-v_{final}^2}{2*a}|$
    \STATE $t_{est}\gets $ time spent traveling at $v_{ego}$ + acceleration time
    \STATE \quad\quad\quad\quad  $=\frac{max(d-d_{accel},0)}{v_{ego}}+\frac{2*d_{accel}}{v_{ego}+v_{final}}$
\ELSE 
    \STATE assume that vehicle accelerates to an intermediate speed $v_i$ before it has to decelerate to reach $v_{target}$
    \STATE $v_i=min(\sqrt{\frac{1}{2}(2*a*d+v_{ego}^2+v_{target}^2)},v_{max})$
    \STATE $d_{accel}=|\frac{v_{ego}^2-v_i^2}{2*a} |+|\frac{v_i^2-v_{target}^2}{2*a}| $
    \STATE $t_{est}=|\frac{v_{max}-v_{ego}}{a}|+|\frac{v_{max}-v_{target}}{a}| +\frac{max(d-d_{accel},0)}{v_{max}}$
\ENDIF
\STATE return $t_{est}$
\end{algorithmic}
\end{algorithm}
The decomposed travel times $t_i$ can then be calculated as follows:
\begin{equation}
    t_i=
    \begin{cases}
        \mathbf{est\_tt}(v_{start}, a_i, v_1, v_{max}, a, d_i) & i=0\\
        \mathbf{est\_tt}(v_i, a_i, v_{end}, v_{max}, a, d_i) &i=n\\
        \mathbf{est\_tt}(v_i, a_i, v_{i+1}, v_{max}, a, d_i) &\text{otherwise}\\
    \end{cases},
\end{equation}
where $v_i$ is the traveling speeds of vehicle $i$; $a_i$ is the acceleration of vehicle $i$; $v_{max}$ is the maximum possible speed on the edge (Equation \ref{eq:vmax}); $v_{start}$ is the assumed speed at the start of the edge, so the speed limit of the edge is used; $v_{end}$ is the assumed speed at the end of the edge, so $1m/s$ is used since edges are controlled with stop signs; and $a$ is the absolute value of the assumed constant acceleration or deceleration rate (a value of $3 m/s^2$ is used).

The maximum possible speed is different on turning connections than on a through connection or a straight road segment. The value of $v_{max}$ is defined as follows: 
\begin{equation}\label{eq:vmax}
    v_{max}=
    \begin{cases}
        10 m/s & \text{left turn (LT)}\\
        8 m/s & \text{right turn (RT)}\\
        \sqrt{2*a*length} & \text{through (TH)}
    \end{cases},\\
\end{equation}
and note that the maximum speed on left turns and right turns is determined from empirical observations.

The cost function (travel time) of each of the 3 routes shown in Fig. \ref{fig:diamond_ap} is therefore formulated as: 
\begin{equation}
    \begin{aligned}
        C&_{p,A\rightarrow C\rightarrow M\rightarrow B}(t) \\
        &=C_{e,TH}(t) + C_{e,a}(t) + C_{e,TH}(t) + C_{e,b}(t)\\
        &+ C_{e,RT}(t) + C_{e,b}(t) + C_{e,LT}(t); \\
        C&_{p,A\rightarrow N\rightarrow D\rightarrow B}(t) \\
        &=C_{e,RT}(t) + C_{e,b}(t) + C_{e,LT}(t) + C_{e,b}(t) \\
        &+ C_{e,TH}(t) + C_{e,a}(t) + C_{e,TH}(t); \\
        C&_{p,A\rightarrow C\rightarrow D\rightarrow B}(t) \\
        &=C_{e,TH}(t) + C_{e,a}(t) + C_{e,RT}(t) + C_{e,c}(t) \\
        &+ C_{e,LT}(t) + C_{e,a}(t) + C_{e,TH}(t). \\
    \end{aligned}
\end{equation}

Based on the estimated real-time travel time of each route, each user (driver) will then select the the route that results in the lowest cost:
\begin{equation}
    \begin{aligned}
        C_u(t)=\min\{C_p(t)\}=\min\{&C_{p,A\rightarrow C\rightarrow M\rightarrow B}(t),\\ &C_{p,A\rightarrow N\rightarrow D\rightarrow B}(t),\\
        &C_{p,A\rightarrow C\rightarrow D\rightarrow B}(t)\};
    \end{aligned}
\end{equation}

\section{Experiment Setup}
To numerically characterize the existence of the BP in microsimulation, several parameters of the baseline and the added-path scenarios are modified to compare the performance of the networks.

First, different inflow rates of vehicles entering the network from node 1, at 50, 200, 400, 600, 800, 900 veh/hr, are used to simulate the network with varying demand. Increasing the inflow rate results in increased density in the system, allowing us to understand how adding a path changes the ability of the system to process demands.

In addition, different speed limits and edge lengths are used to dial the attractiveness of the added path. The speed limits of the edges are changed from 35 m/s to 15 and 10 m/s, while fixing that of the added path ($C\rightarrow D$) to 35 m/s. The combination of edge length and speed limit determine the maximum allowable speed on the edge given a maximum acceleration. When the allowable speed on the added path is much higher than those on the other edges, the added path becomes more attractive, thus allowing us to explore how varying levels of attractiveness of the added path could impact the onset of the BP.

Lastly, multiple inflow nodes (e.g. Node 2, 3, 5, and 7) are used to add additional demand to the network at once to compare the performance of the networks when the demand reaches near-capacity levels.

\begin{figure}[h]
    \centering
    \subfloat[Gird network with 2 routes with multiple inflow nodes (the baseline network/scenario).
    \label{fig:mf_diamond_ori}]{%
   \includegraphics[width=0.48\linewidth]{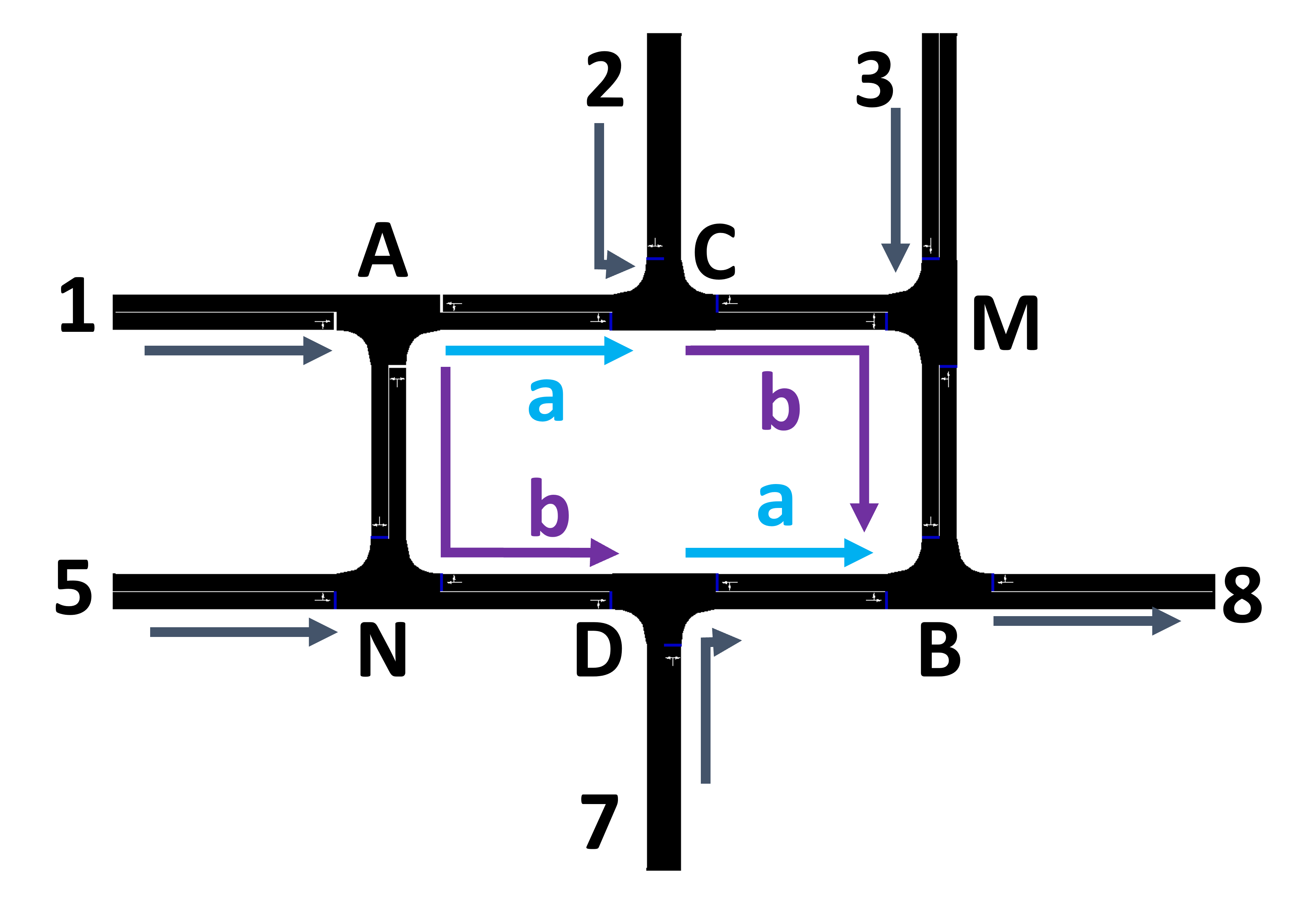}}
   \hfill
   \subfloat[Grid network with an added path and multiple inflow nodes (the add-path network/scenario).
    \label{fig:mf_diamond_ap}]{%
   \includegraphics[width=0.48\linewidth]{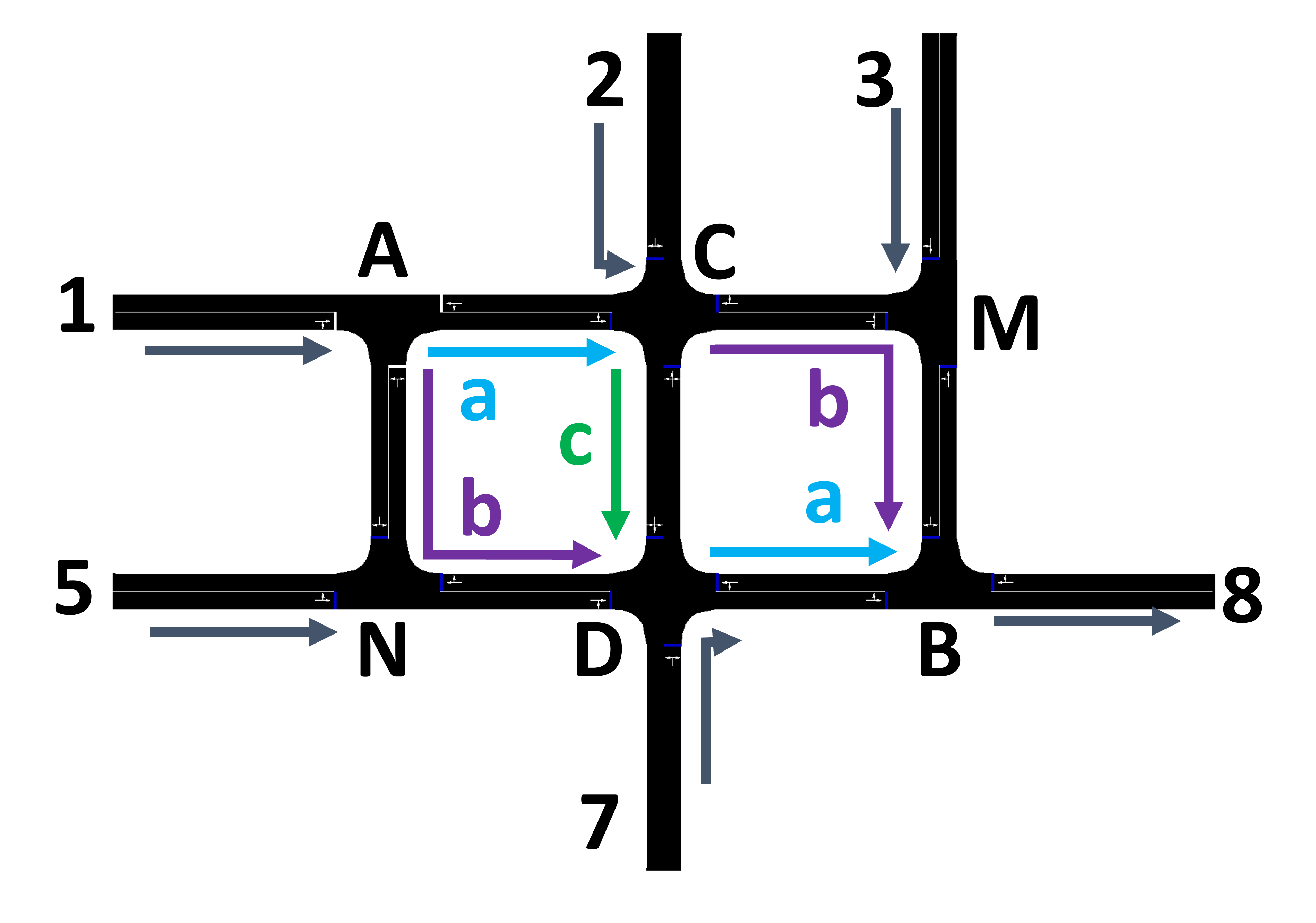}}
  \caption{A 2-by-3 grid network resembling the classical diamond network with multiple inflow nodes.}
  \label{fig:mf_diamond} 
\end{figure}

\section{Results}
\label{sec:results}
We present how adding a path changes the overall vehicle flow, vehicle travel time and capacity of the network compared to the baseline network without any added path.

\subsection{Impact on the Network Flow}
We compare how the output flow rate of networks, given various inflow demand, differs between the baseline scenario and the scenario with an added path.

\begin{figure} 
    \centering
    \subfloat[Output Flow vs. Demand for networks with 50-meter edges.
    \label{fig:outflow_demand_50m}]{%
  \includegraphics[clip,width=0.95\linewidth]{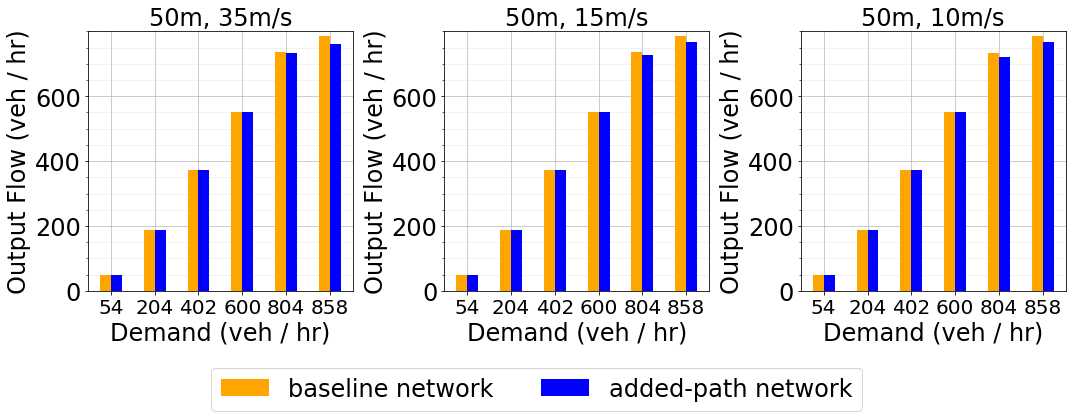}}
   
  \subfloat[Output Flow vs. Demand for networks with 400-meter edges.
    \label{fig:outflow_demand_400m}]{%
  \includegraphics[clip,width=0.95\linewidth]{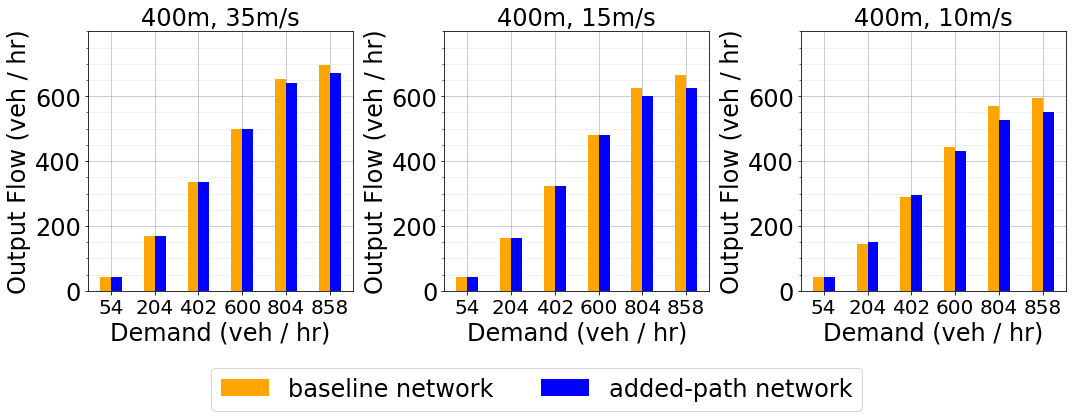}}
  \caption{Comparison of output flow vs. demand for baseline and added-path networks. }
  \label{fig:outflow_demand} 
\end{figure}

Fig. \ref{fig:outflow_demand} shows a comparison of the flow rates of vehicles leaving the network at different demand levels. When the edge length is 50 meters with 35 m/s, 15 m/s, and 10 m/s speed limits (Fig. \ref{fig:outflow_demand_50m}), the output flow rate of the baseline scenario and the added-path scenarios are almost equal at most demand levels. This is because with 50m edges, the maximum achievable speed remains the same due to the limited distance for acceleration and deceleration. Therefore, vehicle throughput is not increased despite the addition of a new route option. 

Moreover, when the edge length is increased to 400 meters (Fig. \ref{fig:outflow_demand_400m}), the output flow of the added-path network becomes even lower as the speed limit on edges other than the added path varies from 35 m/s to 10 m/s. 

The average reduction in output flow rate varied from 2.5\% to 5.7\% as edge length increases from 50m to 400, and from 3.7\% to 5.3\% as the speed limit decreases from 35 m/s to 10 m/s. The scenario with the most attractive added path (400m, 10m/s) shows a 7.4\% reduction in output flow, while with the least attractive added path (50m, 35m/s) a 3\% reduction.

Overall, the analysis shows that adding a path reduces network output flow. This reduction is more significant when the added path is more preferable and demand level is high.

\begin{figure}[h]
    \centering
    {%
   \includegraphics[clip,width=0.95\linewidth]{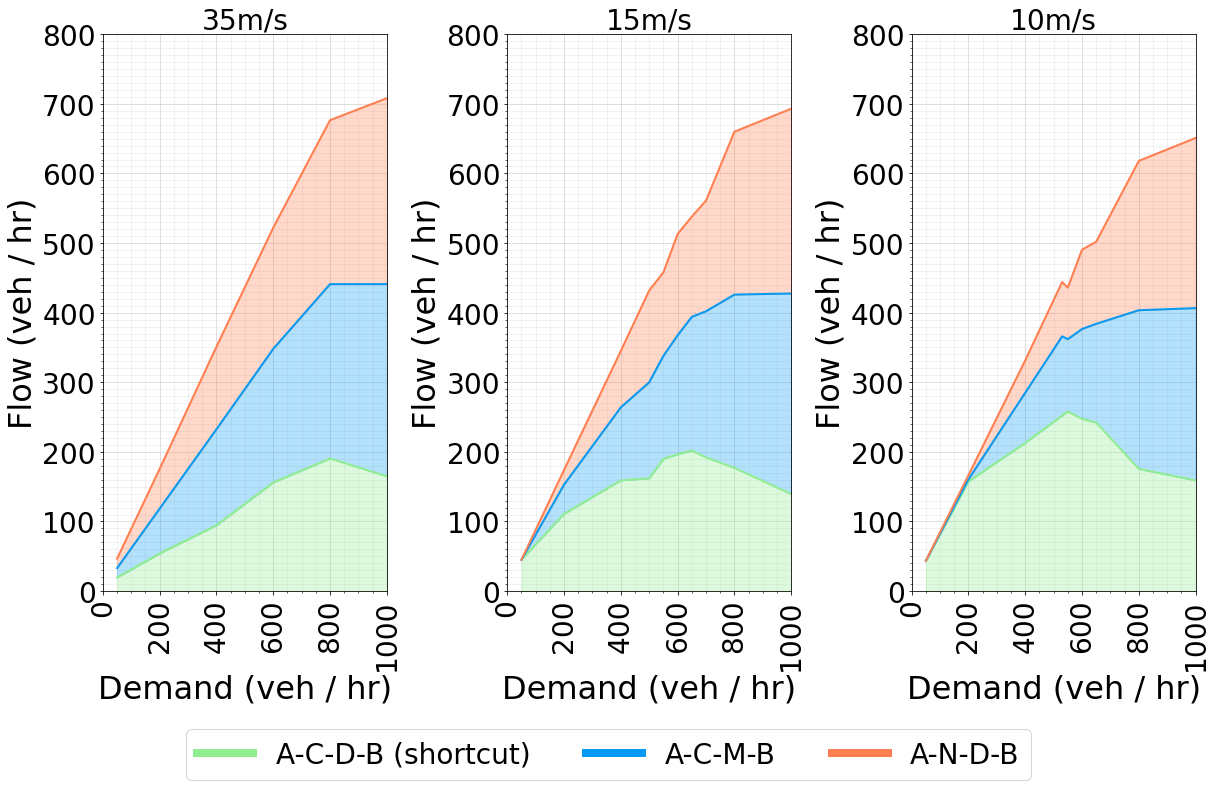}}
  \caption[LoF entry]{Vehicle flow on each route in the added-path network. Each subplot shows the flow per route averaged over different edge length scenarios (50, 200, 300, and 400m) and the same speed limit (35m/s, 15m/s and 10m/s).}
  \label{fig:flow_demand_by_speed} 
\end{figure}

Figure \ref{fig:flow_demand_by_speed} presents the vehicle flow on each route in the added-path network. When the speed limit varies from 35m/s to 10m/s, the added path becomes more attractive. Although flow on each original route increases and converges to a constant as demand level increases, the flow on the shortcut route increases and then decreases at an inflection point. The inflection point occurs at lower demand level as the added path becomes more attractive. 

In Figure \ref{fig:flow_demand_by_speed} the flow rate is slightly higher on $A-C-M-B$ than $A-N-D-B$ with low demand.
In fact, vehicles that choose the shortcut at node $A$ may see a worse shortcut travel time when at node $C$ due to impacts from those already on the shortcut. 
Those that change routing decision between $A$ and $C$ contribute to the higher flow on $A-C-M-B$.

\subsection{Impact on Vehicle Travel Time}
\begin{figure}[h]
    \centering
    {%
   \includegraphics[clip,width=0.95\linewidth]{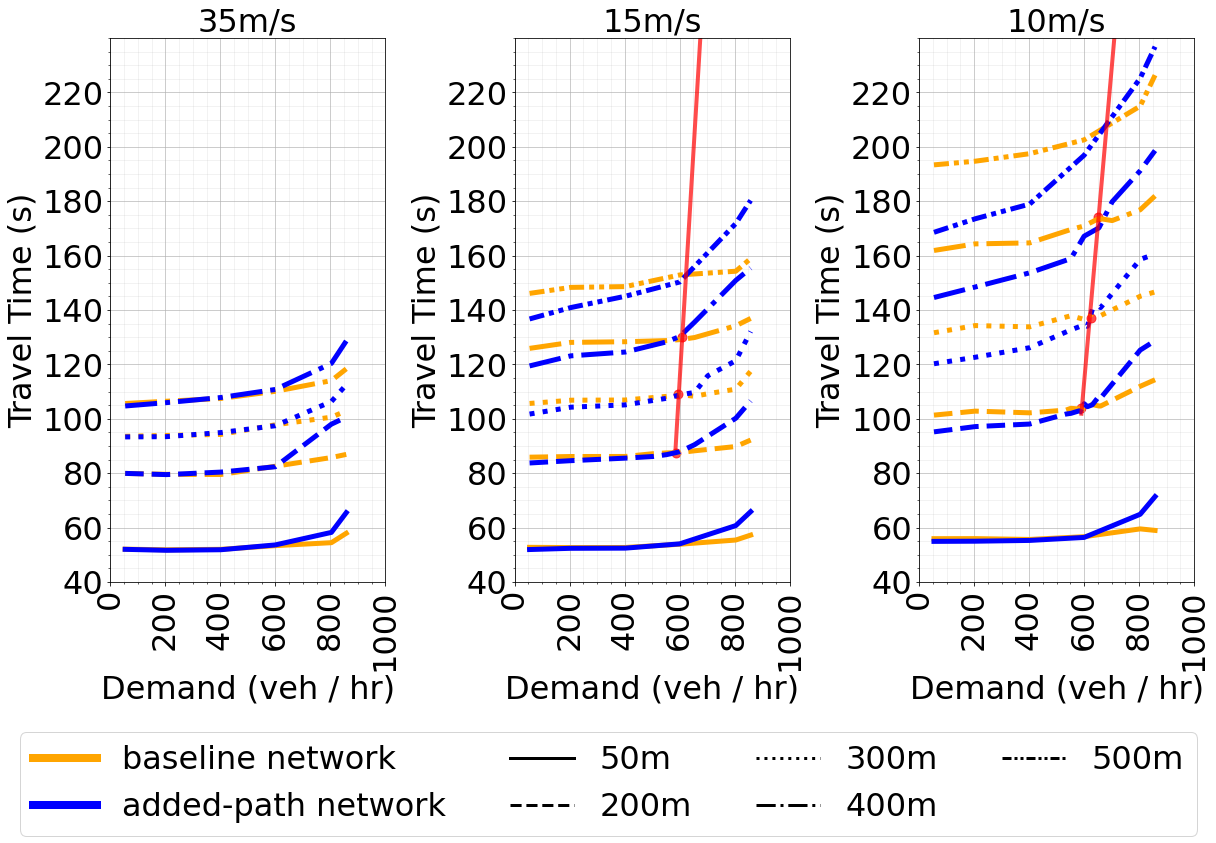}}
  \caption[LoF entry]{Comparison of travel times at different demand level. Crossing points of the lines indicate when the BP appears. An additional experiment with 500m edge is added to test whether it lies on the same line as the other scenarios.}
  \label{fig:tt_demand_all} 
\end{figure}

We also explored the impact of added path to the average travel time from node $1$ to $8$. As shown in Fig. \ref{fig:tt_demand_all}, the average travel time in the added-path network is larger than in the baseline network with high demand. The added shortcut brings advantage to travel times when demand is low. The average increase in travel time varies from 10\% for networks with 400m to 16\% with 50m. The largest increase is 21\% in the network with 50m edges and 10 m/s speed limits, and the smallest increase is 8.6\% with 400m and 35 m/s.

\begin{figure} 
    \centering
    {%
   \includegraphics[clip,width=0.95\linewidth]{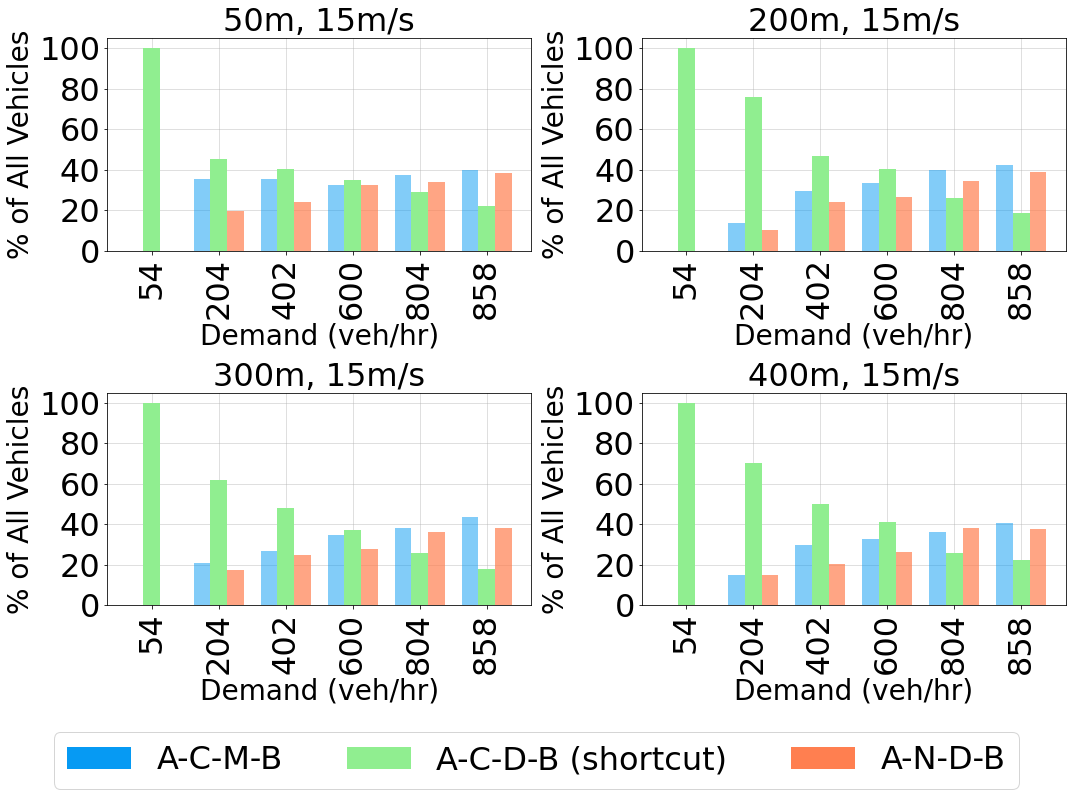}}
  \caption[LoF entry]{Percentage of vehicles using each route with edge lengths of 50, 200, 300, and 400m and 15m/s speed limit.}
  \label{fig:perc_veh_15mps} 
\end{figure}

When demand is low, vehicles almost always choose the shortcut route as shown in Fig. \ref{fig:perc_veh_15mps}. These vehicles complete their trips and give opportunities for future vehicles to choose the shortcut. With high demand the shortcut is occupied faster, so the following vehicles are forced to use the other routes and the advantage of using the shortcut diminishes.

For networks with the same speed limit, the "critical point" (i.e. where the BP shows up) appears at higher demand level with longer edges. Critical points show linearity with the edge length and the demand level. As shown in Fig. \ref{fig:tt_demand_all}, the fitted line from the critical points of 50, 200, 300, and 400m cases cross the critical point of the additional 500m case.

\subsection{Impact on Network Capacity}

\begin{figure} 
    \centering
    {%
  \includegraphics[clip,width=0.9\linewidth]{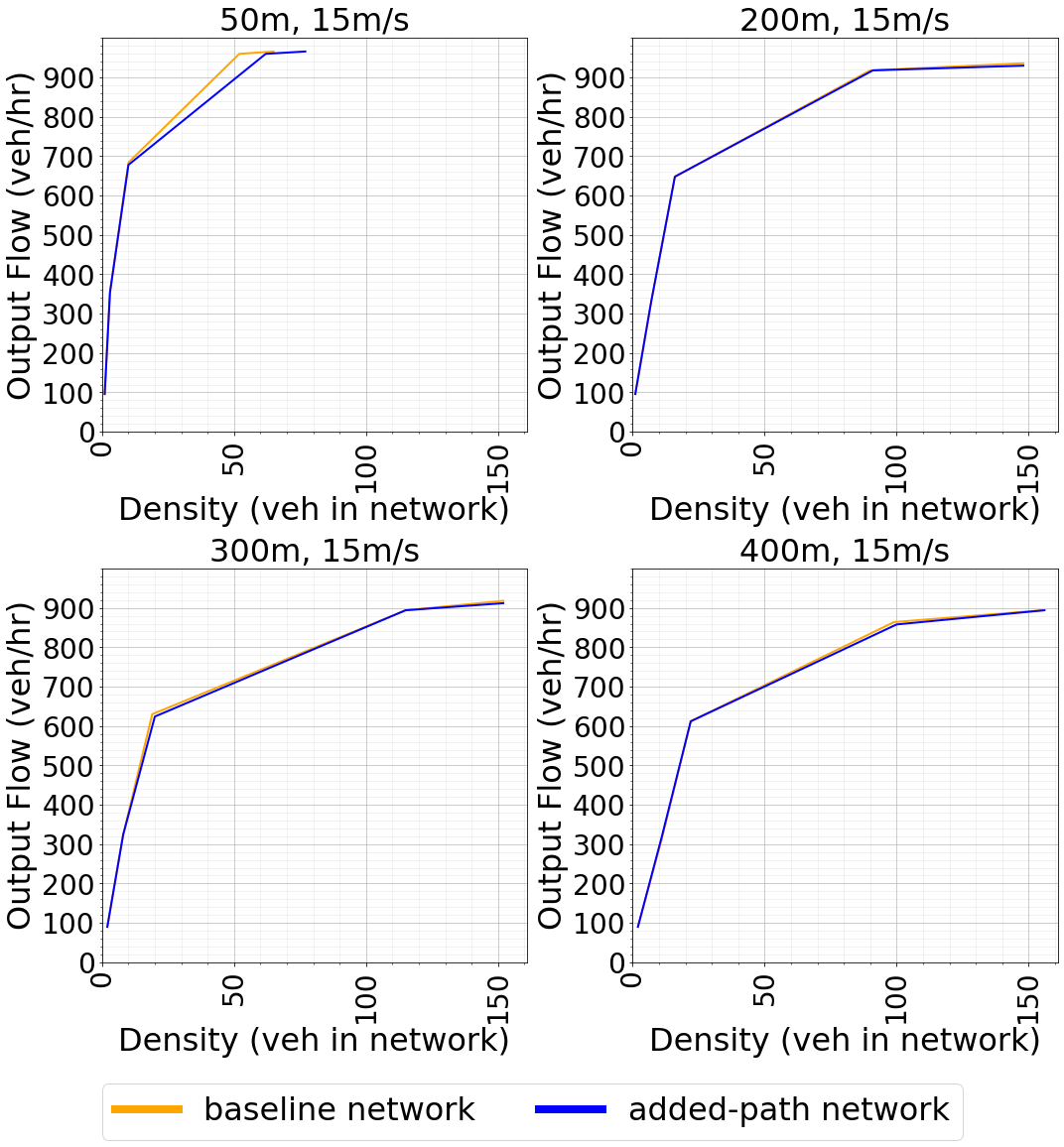}}
  \caption[LoF entry]{Comparison of flow-density curves of both types of networks with different edge lengths and a speed limit of 15m/s on edges other than the added path.}
  \label{fig:fd_mf} 
\end{figure}

We explore letting vehicles enter at Node 2, 3, 5, 7, in addition to Node 1, and plot the flow-density curves of the networks to compare how the capacity of the baseline and added-path networks differs. As shown in Fig. \ref{fig:fd_mf}, although adding a path brings more physical space, the actual capacity of the added-path network is almost the same as that of the baseline network in all scenarios.


Simulation reveals that the attractiveness of the shortcut route results in the downstream edge getting filled up faster than other edges, making the shortcut route undesirable. Therefore, though vehicles choose the shortcut route at first, the segments upstream and downstream of the added path becomes congested over time, making it harder to get to the added path, resulting in the added path being under-utilized.

\section{Conclusion and Discussion}
\label{sec:conclusion}
In this paper, we have designed a grid network with and without an added path to explore the Braess's Paradox in the dynamic microsimulation environment. Different from existing research about BP that assigns traffic on possible routes instantaneously, we allow each vehicle to dynamically choose the route with the lowest estimated travel time. By varying the edge length and speed limit, we discuss the impact of the added path on network flow, vehicle travel time, and network capacity. We find that adding a path to the 2-by-3 grid network does not increase network output flow in any experiment scenarios, and reduces the output flow by as much as 7.4\% when the added path is more attractive. Traffic flow on the added path increases and then decreases after a inflection point as demand level increases. In all experiment scenarios, the added path result in increased average vehicle travel times by as much as 21\% after a critical demand level. Moreover, the increased congestion resulted from vehicles using the added path makes the additional physical room brought by the added path underutilized.

We could further explore how different parameters of the infrastructure affect the network performance more precisely. Different from dynamic traffic assignment related models, a microscopic perspective provides a more advanced view on the BP, in which the impact of an added path can be characterized more finely in terms of the network characteristics. 
As such, this study provides practical examples that adding new capacity to the road network may not increase operational capacity and even worsen travel times. A systematic review of other networks using the same approach could help generalize the conclusion more generically. In the future, we would like to 1) give sensitivity analysis to the design of user travel cost function; 2) explore how real-world traffic networks might suffer from the BP; 3) consider how control strategies may bring a difference. We are also interested to consider the theoretical and conceptual differences between the static and dynamic formulations.

\bibliographystyle{IEEEtran}
\bibliography{IEEEabrv,ref}
\end{document}